# Barriers to Digital Mental Health Services among College Students

Ha Na Cho[a], Kyuha Jung[a], Daniel Eisenberg[b], Cheryl A. King[c], Kai Zheng[a]
[a] *University of California, Irvine*
[b] *University of California, Los Angeles*
[c] *University of Michigan, Michigan*

**Abstract.** This qualitative study explores barriers to utilization of digital mental health Intervention (DMHI) among college students. Data are from a large randomized clinical trial of an intervention, *e*Bridge, that used motivational interviewing for online counseling to connect students with mental health issues to professional services. We applied thematic analysis to analyze the feedback from the student participants regarding their experience of using the DMHI platform. We identified nine key barriers to DMHI adoption and the use of in-person mental health services: emotional distress, time constraints, privacy concerns, resource accessibility, financial challenges, medication stigma, dissatisfaction with communication, content clarity, and treatment-related concerns. Our findings emphasize the need for personalized, culturally sensitive interventions and improved strategies to enhance the access and engagement in mental health support for young adults.

## 1. Introduction

In recent decades, suicide rates in the United States have risen by approximately 19.57% from 2011 to 2024 [1]. Among college students, suicide is the second leading cause of death, with rates increasing across all racial and ethnic adolescent groups [2]. Despite this, less than 27% of college students with mental health (MH) issues seek and receive appropriate treatment [3]. Digital mental health interventions (DMHIs) are crucial for supporting behavioral changes and providing ongoing accountability, particularly in young adults facing barriers to in-person care, such as mobility challenges or financial constraints [4]. Studies show that DMHIs effectively reduce anxiety and are preferred by many students, with significant numbers choosing online counseling over traditional methods [5]. However, dropout rates range from 2% to 83%, highlighting the need for improved engagement strategies [6].

The *e*Bridge study, a randomized controlled trial across four U.S. universities, evaluated a brief intervention for college students at risk for suicide [7]. Participant received either a list MH resources (control) or additional personalized feedback with brief motivational interviewing via an online portal (intervention). The intervention engagement remained modest with 21.2% logging in at least once and only 11.2% posting a message, underscoring the challenges of sustaining student engagement.

This study aims to analyze qualitative data from the *e*Bridge project to explore barriers to engagement with online MH services among college students [8]. Data were collected from participant interactions with counselors, including counselor's feedback on open-ended survey responses. The objective was to identify behavioral factors, such as students' use of the online intervention, and attitudinal factors, reflecting their perceptions of its effectiveness, that influence engagement decisions. Insights from this analysis can guide improve technology design or more effective strategies to enhance DMHI utilization among college students.

## 2. Methods

Participants included students from four universities in the U.S. who met eligibility criteria: 18 years or older, enrolled in a degree-seeking program with more than one semester remaining, not be currently receiving MH counseling, and identified as at risk for suicide. The specific criteria are detailed in previous reports [8]. Intervention group participants accessed a secure portal for asynchronous text communication with trained MH counselors, who provided encouragement and treatment information but not therapy. Participants later provided feedback on perceived barriers to engaging with the intervention.

The qualitative data analyzed in this study were from two sources: (1) online chat records between counselors and students, and (2) free-text responses from the post-intervention open-ended survey questions on participants' experiences. All data were de-identified to protect privacy, with names replaced by unique study IDs. Thematic analysis [9] was applied to transcripts and survey responses using ATLAS.ti version 23.3.0 for Mac [10]. The process involved familiarization, exploration, and open coding. Themes and subthemes were refined by merging similar codes and eliminating redundancies. Two coders (HC, KJ) collaborated to validate and finalize the themes.

## 3. Results

### 3.1. Participant Demographics

From 2015 to 2018, 178,879 students were invited to participate in the larger eBridge study, with 40,347 (22.5%) completing the initial screening survey. Of these, 14.4% were identified with suicide risk, further narrowed to 8.1% after excluding those already receiving MH treatment. Half of the eligible group were randomized to the intervention, yielding in a sample of 1,673 students. The intervention group consisted of 61.6% female and 38.4% male participants. Detailed information on demographics, race, and age distribution, is provided in Table 1.

**Table 1.** Participants' characteristics.

| Parameters | Number (N= 1,673) | Percentage (%) |
|---|---|---|
| Gender | | |
| Female | 1,042 | 62.3 |
| Male | 580 | 34.6 |
| Transgender/genderqueer | 51 | 3.0 |
| Age | | |
| 18 | 652 | 39.0 |
| 19-22 | 642 | 38.4 |
| 23-30 | 325 | 19.4 |
| 31+ | 52 | 3.1 |
| Race | | |
| White | 1,205 | 72.0 |
| Black | 124 | 7.4 |
| Asian | 351 | 21.0 |
| Other (Indian, Hispanic) | 209 | 12.5 |

*3.2. Barriers to Digital Mental Health Adoption*

*Theme 1. Communication: Lack of Human Interaction:* Some participants noted the lack of direct communication in online text-based interventions, doubting their effectiveness compared to in-person interactions. One participant noted: "*I was looking for something a little more personal and direct, which is hard to provide solely through text…this felt impersonal and even insincere*" (P.736). The lack of empathy in digital formats made it easy to dismiss online suggestions: "*It is easy to just close the laptop screen and continue on with your day ignoring it if you are not ready to change*" (P.18174). Another participant reflected, "*That voice was coming from a computer screen…the lack of face-to-face contact was not as appealing for me*" (P.19218).

*Theme 2. Digital content: Lack of Clarity, Engagement, and Personalization:* Many found multiple-choice options in the screening questions limiting and preferred more opportunities to elaborate: "*The multiple-choice options are limiting…I prefer elaborating on my response*" (P.23924). The content lacked necessary personalization, often feeling generic and impersonal, which reduced participants' engagement. Additionally, digital feedback was challenging to comprehend due to formatting issues: "*There is just a lot of text on the screen*" (P.32) and "*hard to read*" (P.6246). The lack of depth and personalization led to a sense of repetitiveness, making it difficult for participants to fully engage with or understand the content.

*Theme 3. Trust and Privacy: Accuracy and confidentiality concerns:* The absence of in-person interaction raised doubts about accuracy: "*I didn't think it was very accurate because they don't actually know me*" (P.156). Establishing trust online was challenging, as participants questioned confidentiality and felt hesitant to fully disclose their feelings: "*Are this information confidential?*" (P.1224) and "*difficult to completely trust what it is saying just by reading the passage*" (P.918). Additionally, concerns about potential career impacts due to mental health records made some participants reluctant to be fully honest, fearing negative consequences: "*The legal issues associated with mental health treatment are far too risky for my goals*" (P.1394).

*3.3. Barriers to in-person Mental Health Intervention*

*Theme 4. Time and scheduling constraints:* Many reported that demanding school schedules prevented them from participating: "*I chose not to participate because I didn't have the time*" (P.21508). Extended waitlists also discouraged engagement: "*The wait is too long*" (P.455). Even after matching with counselors, appointment times often did not align with participants' schedules, causing frustration and delays in accessing support. Limited facility availability further hindered timely MH services: "*I was told they were completely booked through the end of the semester*" (P.301), "*I tried to set up an appointment…I have not gotten a response email.*" (P.566).

*Theme 5. Financial issues: Cost barrier:* Anxiety over high costs deterred many participants, despite their desire for care: "*I've considered going to a psychiatrist, but that can be expensive*" (P.69). Issues with insurance coverage added to these barriers: "*My insurance doesn't cover mental healthcare*" (P.240). Participants were also worried about family members learning of their MH status through insurance claims: "*I don't want my parents to know*" (P.5). Financial concerns were often met with inadequate support, leading to discouragement: "*When I expressed concern about the out-of-pocket cost…the counselor asked me why I was still pursuing it. This made me feel discouraged*" (P.5656).

*Theme 6. Treatment-related concerns:* Participants feared that medication would worsen symptoms: "*I always end up on medication which makes things worse*" (P.1208) or were apprehensive about being prescribed medication: "*I'm concerned…they may want to try and put me on medication*" (P.1110). Fear of social stigma also influenced participants, with some reporting family pressure to discontinue medication: "*I was prescribed medications…but my parents asked me to put an end to it*" (P.793). Additionally, some participants doubted the efficacy of medications: "*Medications have not worked for me*" (P.1093) or generally disliked taking them.

*Theme 7. Lack of resource accessibility:* Participants highlighted a lack of guidance and limited resources, which made navigating MH options challenging. This led some to opt for less effective alternatives, such as educational workshops: "*The referral to resources seemed a bit hard to navigate and may have contributed to my choice to seek more educational workshops than professional counseling*" (P.15162). Many participants were uncertain about available resources and expressed difficulties accessing interventions due to cumbersome procedures: "*It's hard when you struggle with anxiety to physically make the call and show up to an appointment*" (P.583). A lack of clear guidance on resources discouraged participants from seeking care.

*3.4. Barriers to both in-person and digital services.*

*Theme 8. Personal reflections: Emotional distress, social judgment, self-criticism, and self-relying:* Emotional distress—such as stress, fear, and anxiety—was often heightened by peer comparisons. Social judgment led to feelings of blame and discouragement, with participants expressing fear of being judged: "*The person may feel attacked*" (P.4178). Many found counselor feedback overwhelming and potentially triggering: "*It was a little overwhelming*" (P.6244). Fear of judgment, doubts about the effectiveness of counseling, and embarrassment were common barriers: "*I am very skeptical about going to counseling sessions and feel embarrassed asking for help*" (P.610). Participants also hesitated due to uncertainty about intervention effectiveness and past negative experiences, opting instead for self-reliance through personal coping strategies like time management.

*Theme 9. Interaction with the counselor: Concerns on approach, fit, and past experiences:* Participants expressed concerns with directive advice and a lack of empathy: "*I don't yet feel like I have found anyone who really wants to help*" (P.8056). Cultural differences also hindered emotional connections, and participants feared that their problems might be dismissed: "*A counselor would say I am overreacting*" (P.1122). Finding a culturally compatible counselors was crucial: "*I would like to speak with another person of color or queer person*" (P.1053). Poor communication led to frustration, highlighting the need for a more empathetic, personalized approach.

## 4. Discussion and Conclusion

This study explores barriers young adults face when engaging with digital and in-person MH services. Thematic analysis revealed nine major barriers: emotional distress, time constraints, counselor issues, resource accessibility, financial challenges, medication concerns, privacy, and dissatisfaction with digital content and human interaction.

Participants' hesitancy toward DMHI highlights the importance of human connection, personalized feedback, and real-time content, which AI-driven agent interventions could enhance. Privacy concerns, especially regarding career impacts, and the absence of in-person empathy highlighted the importance of identity-sensitive care. Some participants preferred self-guided interventions, suggesting conversational agents could support self-management if ethical considerations are upheld [11]. Barriers to in-person interventions, including time constraints, MH provider shortages, and response delays, underscored the need for solutions like online personalized scheduling systems. Financial challenges such as costs and insurance coverage emphasized the importance of affordable MH services through insurance or government-funded grants. Concerns about medication side effects and stigma stressed the need for shared decision-making, education, and public awareness campaigns to destigmatizing MH treatment. Furthermore, emotional distress, social judgment, and skepticism reduced engagement, highlighting the need for culturally sensitive interventions. Self-stigma and fear of vulnerability led participants toward self-management over external support. Co-designed interventions can help reduce stigma and address personalization needs. Participants valued empathetic, culturally competent care, flexible formats (e.g., video, or non-video), and long-term support for meaningful MH care.

Our findings emphasize the need for a tailored approach to enhance young adults' engagement in DMHI. Refining DMHI platforms and adopting participant-centered strategies can reduce barriers and improve support.